\documentclass[runningheads]{llncs}
\usepackage[T1]{fontenc}
\usepackage{amssymb,amsmath,graphicx}
\usepackage{color}
\usepackage[margin=1in,footskip=0.25in]{geometry}

\usepackage{placeins}
\usepackage{url}
\usepackage{hyperref}
\hypersetup{colorlinks,allcolors=blue}
\usepackage{babel}
\usepackage{caption}
\usepackage{subcaption}

\begin{document}
\title{Certified Lipschitz Bounds for Hybrid Quantum-Classical Decision Models}
\titlerunning{Certified Lipschitz Bounds for Hybrid Models}
%
\author{
Mohammad Saeed Arvenaghi\inst{1}
\and
Sajjad Hashemian\inst{2}
}
\authorrunning{M. S. Arvenaghi, S. Hashemian}
%

\institute{
Iran University of Science and Technology, Tehran, Iran\\
\email{m\_arvenaghi@mathdep.iust.ac.ir} \and
University of Tehran, Tehran, Iran\\
\email{sajjadhashemian@ut.ac.ir}
}
\maketitle              

\begin{abstract}
We study certified upper bounds on Lipschitz constants for hybrid quantum-classical decision models, which combine classical neural-network layers with parameterized quantum circuits. We pair an SDP-based certificate for classical layers with an exact spectral characterization for probability-output quantum models. For hybrid architectures, these componentwise estimates are connected through the encoding maps to obtain a certified composition bound for the complete model. The resulting bounds support sensitivity analyses relevant to robustness, stability, generalization, and fairness.

\end{abstract}

\section{Introduction}

Hybrid quantum-classical decision models interleave classical neural-network components with parameterized quantum circuits, with applications ranging from transfer learning and classification to other data-driven decision tasks \cite{Mari2020transferlearningin,schetakis2022review}.

In such architectures, information repeatedly passes between classical vectors and quantum states.
As a result, sensitivity analysis must account not only for the Lipschitz behavior of the individual components but also for the interfaces that connect them.
Lipschitz bounds are widely used to quantify sensitivity to input perturbations and to support robustness, stability, generalization, and fairness analyses.
For classical neural networks, Fazlyab \emph{et al.} \cite{fazlyab2019efficient} developed LipSDP, an SDP-based method for computing certified upper bounds.
In the quantum setting, Guan \emph{et al.} \cite{guan2022fairness} characterized the Lipschitz constant of probability-output quantum decision models with trace distance at the input and total variation distance at the output.
More recently, Berberich \emph{et al.} \cite{berberich2024training} derived Lipschitz bounds for quantum models and used them for regularized training, emphasizing the role of the data encoding.
These results provide useful certificates for individual classical or quantum components, but they are expressed with different domains, codomains, and metrics.

This mismatch becomes important in a hybrid model.
Classical blocks are naturally analyzed in Euclidean norm, an encoding map sends a classical vector to a density operator measured in trace distance, a probability-output quantum model is naturally measured in total variation distance, and an expectation-value quantum layer returns a Euclidean vector.
Therefore, a componentwise bound can be propagated through the complete architecture only after the encoding maps and the metric conversions at the classical--quantum interfaces are made explicit.
Our focus is this interface problem: constructing a type-consistent end-to-end certificate from componentwise bounds without interpreting the hybrid model as a single SDP.

We have three main contributions in this work.
First, we formulate a metric-consistent composition framework for hybrid quantum-classical models that explicitly includes the encoding maps.
The framework combines an existing classical LipSDP certificate with an existing exact characterization for probability-output quantum decision models and yields a certified upper bound for the complete hybrid architecture.
Second, for the expectation-value output used in practical variational quantum models, we give a trace-distance-to-Euclidean Lipschitz bound and specialize it to local Pauli-$Z$ observables with product $R_X$ encoding; for an $m$-qubit block, this gives the explicit quantum-block factor $\sqrt{m}$.
Third, we implement the resulting end-to-end certificate in a hybrid classifier and study certificate-aligned Lipschitz regularization together with naive and PGD training.
The experiments show how the certified upper bound changes with the training strategy and expose the trade-off between sensitivity reduction and classification accuracy when the regularization is made too strong.

The resulting guarantee is deliberately componentwise: it does not claim that the global Lipschitz constant is obtained by a single convex optimization problem or that the composition bound is tight.
Section~2 introduces the classical, quantum, and hybrid models.
Section~3 develops the componentwise certificates and their metric-consistent composition, and Section~4 reports the corrected numerical study.

\section{Preliminaries}
\subsection{Classical Neural Networks}
At a high level, a \emph{classical neural network} is a parameterized function that consists of an input layer, one or more hidden layers, and an output layer. Each layer applies a sequence of linear transformations followed by a non-linear activation function, thus enabling the network to capture intricate non-linear relationships in the data. A popular choice for activation functions is the Rectified Linear Unit (ReLU), which is Lipschitz continuous with a constant of 1. Other activations such as the sigmoid and hyperbolic tangent functions are respectively $1/4$-Lipschitz and $1$-Lipschitz. However, while individual layers may exhibit favorable Lipschitz properties, the overall Lipschitz constant of a deep network can grow exponentially with the depth if not properly controlled \cite{cisse2017parseval}\cite{gouk2021regularisation}. 

The learning process involves adjusting the weights and biases of each layer by minimizing a loss function—typically via gradient descent or its variants \cite{bottou2010large}. 
Despite the empirical success of these methods, controlling the Lipschitz constant of the network has emerged as a central theme in understanding the robustness, generalization, and fairness of the learned models\cite{gupta2021individual}. In classical learning theory, the Lipschitz constant has a long history of study; early works by Bartlett \emph{et al.} \cite{bartlett2017spectrally} and Neyshabur \emph{et al.} \cite{neyshabur2018the} highlighted its role in bounding the generalization error. More recently, its importance in mitigating adversarial vulnerability and ensuring fairness has been emphasized \cite{dwork2012fairness}.

To address these issues, recent research has focused on explicitly constraining or regularizing the Lipschitz constant during training. By imposing Lipschitz constraints, one can design networks that are more robust to input perturbations, thereby enhancing their generalizability and fairness. For example, methods based on semidefinite programming \cite{fazlyab2019efficient} and spectral normalization \cite{miyato2018spectral} have been proposed to achieve tighter control over the Lipschitz constant. Such approaches not only improve the stability of the training process but also provide theoretical guarantees that are crucial for applications in safety-critical systems.

\begin{definition}[Neural Network]
\label{nn}
A \emph{classical neural network} $\mathcal{C}$ is defined as a composition of $d$ layers $\{L_{n_{i-1}\to n_i}\}_{i=1}^d$, where each layer is a mapping
\[
L_{n_{i-1}\to n_i}(x) = \sigma(W_i x + b_i),
\]
with $W_i\in\mathbb{R}^{n_i\times n_{i-1}}$, $b_i\in\mathbb{R}^{n_i}$, and $\sigma:\mathbb{R}\to\mathbb{R}$ a Lipschitz continuous non-linear activation function (e.g., ReLU, sigmoid, or tanh). The network is then given by the composition
\[
\mathcal{C}(x)=L_{n_{d-1}\to n_d}\big(L_{n_{d-2}\to n_{d-1}}(\cdots L_{n_0\to n_1}(x)\cdots)\big),
\]
where $n_0, n_1,\dots,n_d$ denote the respective layer dimensions.
\end{definition}

This formulation encapsulates the fundamental structure of neural networks, providing a basis for analyzing their robustness and generalization properties via the Lipschitz constant. In summary, while ensuring that each layer is Lipschitz continuous is a necessary condition for robust behavior, it is not sufficient on its own; the aggregate effect across layers must be carefully managed.

\subsection{Quantum Decision Models}
Quantum decision models utilize the principles of quantum mechanics to process information and decision-making. In these models, the data is encoded into quantum states, which are then manipulated by parameterized quantum circuits often referred to as quantum variational circuits. 

A quantum variational circuit can be viewed as a parameterized map acting on density operators. Let $\mathcal{H}$ be a finite-dimensional Hilbert space and denote by $\mathcal{D}(\mathcal{H})$ the set of density operators (i.e., positive semidefinite operators with unit trace) over $\mathcal{H}$. A typical quantum decision model outputs a probability distribution over a set of measurement outcomes. This is achieved by evolving the input state using a sequence of unitary operations and subsequently performing a quantum measurement. The inherent probabilistic nature of measurements and the high-dimensional structure of Hilbert spaces provide a framework for modeling complex decision boundaries that may be challenging for classical systems \cite{schetakis2022review}\cite{liang2020variational}.

\begin{definition}[Quantum Variational Circuit]
A \emph{quantum variational circuit} is a function 
\[
\mathcal{C}: \mathcal{D}(\mathcal{H}) \to \Delta^{|\mathcal{O}|-1} \subset \mathbb{R}^{|\mathcal{O}|},\quad
\rho\mapsto\left\{ \mathrm{tr}\left(M_i\,\mathcal{E}(\rho)\,M_i^\dagger\right) \right\}_{i\in \mathcal{O}}
\]
where $\rho\in\mathcal{D}(\mathcal{H})$ is the input quantum state. $\mathcal{E}=\prod_{i=1}^{m} \mathcal{E}_i$ is a unitary, completely positive and trace-preserving map representing the evolution of the quantum state, composed of a sequence of unitary gates $\{\mathcal{E}_i\}_{i=1}^m$. $\{M_i\}_{i\in\mathcal{O}}$ is a collection of measurement operators forming a positive operator-valued measure (POVM) on $\mathcal{H}$, with $\mathcal{O}$ indexing the set of measurement outcomes, and the output $\mathcal{C}(\rho)$ is a probability vector in the simplex $\Delta^{|\mathcal{O}|-1}$, with each component corresponding to the probability of obtaining the respective measurement outcome.
\end{definition}
In this setup, the role of each quantum gate $\mathcal{E}_i$ in $\mathcal{C}$ is analogous to the layers in a classical neural network, where each layer transforms the data and propagates it forward. The measurement operators $\{M_i\}$ perform the final stage of the computation, converting quantum information into classical data that can be interpreted as decision outcomes.
In classical settings, Lipschitz continuity is a well-established concept used to control the sensitivity of functions to input perturbations. In the quantum realm, analogous techniques have been developed to bound the variation in the output probability distributions with respect to perturbations in the input state. Guan \emph{et al.} \cite{guan2022fairness} elegantly utilize the quantum operators to estimate the Lipschitz constants of quantum variational circuits. This methods involve comparing the trace distance between quantum states with the total variation distance between the corresponding measurement outcomes.

\subsection{Hybrid Quantum-Classical Decision Models}
Hybrid quantum-classical decision models integrate the classical neural networks with the capabilities of quantum variational circuits. Such architectures seek to exploit the complementary strengths of both paradigms: classical networks offer robust, scalable architectures with mature training methods, while quantum circuits can, in principle, capture and process high-dimensional correlations and quantum effects beyond classical simulation capabilities \cite{Mari2020transferlearningin}.

In a typical hybrid architecture, the overall model is expressed as a composition of classical layers interleaved with quantum subroutines. 

\begin{definition}[Hybrid Quantum-Classical Neural Network]
\label{hybridnn}
A \emph{hybrid quantum-classical neural network} is a parameterized function
\[
f \colon \mathbb{R}^{d_0} \to \mathbb{R}^{d_{k+1}},\qquad
f = f_{k+1}\circ q_k\circ E_k\circ f_k\circ\cdots\circ q_1\circ E_1\circ f_1,
\]
where, for each \(i=1,\dots,k\),
\[
f_i:\mathbb{R}^{d_{i-1}}\to\mathbb{R}^{m_i},\qquad
E_i:\mathbb{R}^{m_i}\to\mathcal{D}(\mathcal{H}_i),\qquad
q_i:\mathcal{D}(\mathcal{H}_i)\to\mathbb{R}^{d_i},
\]
and \(f_{k+1}:\mathbb{R}^{d_k}\to\mathbb{R}^{d_{k+1}}\).
Thus, each classical output is first encoded as a quantum state before it is passed to the corresponding quantum component.
The map \(q_i\) may return a measurement-probability vector, as in Theorem~\ref{thm:lipconstant}, or an expectation-value vector equipped with a certified trace-distance-to-Euclidean bound as used in Section~\ref{sec:experiments}.
\end{definition}

For each $i=1,\dots,k$, the encoding map $E_i$ is an explicit component of the hybrid composition.
For the Lipschitz analysis below, we assume that it admits a bound $K_{E_i}$ satisfying $D(E_i(u),E_i(v))\le K_{E_i}\|u-v\|_2$.
Common encoding schemes include amplitude encoding, angle encoding, and basis encoding, each with distinct trade-offs in terms of circuit depth and error robustness \cite{schetakis2022review}.
This modular structure permits a componentwise Lipschitz analysis, provided that the metrics at the classical--quantum interfaces and the encoding maps are treated explicitly.

\section{Lipschitz Constant Estimation}
The Lipschitz constant of a function is a fundamental quantity in both classical and quantum learning theory, as it quantifies the sensitivity of the function's output to perturbations of input. A function \( f \) is said to be Lipschitz continuous if there exists a constant \( K \ge 0 \) such that for all inputs \( x,y \) in its domain,
\[
\| f(x)-f(y) \| \le K \|x-y\|.
\]
The smallest value of such a constant $K$ is denoted by $K^*$ and is called the Lipschitz constant of the function $f$.
In the context of deep neural networks, controlling the Lipschitz constant has been pivotal in establishing robustness, generalization, and fairness properties. 
In this section, we review the SDP certificate used for classical neural networks, state the exact spectral characterization for probability-output quantum decision models, and combine componentwise bounds to certify the complete hybrid model.

\subsection{Lipschitz Estimation in Classical Neural Networks}

A major line of work has focused on semidefinite programming (SDP) certificates for upper bounds on the Lipschitz constants of feed-forward neural networks. In \cite{fazlyab2019efficient}, the authors derive an SDP formulation based on the observation that many popular activation functions (e.g., ReLU, sigmoid, tanh) are not only Lipschitz continuous but also satisfy incremental quadratic constraints as a consequence of being gradients of convex potential functions. This observation leads to the following result.

\begin{proposition}[LipSDP \cite{fazlyab2019efficient}]
\label{thm:lipschitz_certificate}
Consider the standard LipSDP setting of \cite{fazlyab2019efficient}.
Let
\[
\mathbf{x}=(x_0^\top,x_1^\top,\ldots,x_L^\top)^\top
\]
denote the stacked input and hidden-layer variables, and write the network compactly as
\[
B\mathbf{x}=\phi(A\mathbf{x}+b),\qquad
f(x)=C\mathbf{x}+b_L,
\]
where \(A\), \(B\), and \(C\) are the block matrices induced by the network weights.
Let
\(E_0=[\,I_{n_0}\;0\;\cdots\;0\,]\), so that \(E_0\mathbf{x}=x_0\).
Suppose the componentwise activation is slope-restricted on \([\alpha,\beta]\), and let \(T\) belong to the multiplier cone \(\mathcal{T}\) defined by the incremental quadratic constraints in \cite{fazlyab2019efficient}.
Define
\[
Q(T)=
\begin{pmatrix}
-2\alpha\beta T & (\alpha+\beta)T\\
(\alpha+\beta)T & -2T
\end{pmatrix}.
\]
If there exist \(\rho>0\) and \(T\in\mathcal{T}\) such that
\begin{equation*}
\mathcal{M}(\rho,T):=
\begin{pmatrix}A\\B\end{pmatrix}^{\!\top}
Q(T)
\begin{pmatrix}A\\B\end{pmatrix}
+C^\top C-\rho E_0^\top E_0
\preceq 0,
\end{equation*}
then
\[
\|f(x)-f(y)\|_2\le \sqrt{\rho}\,\|x-y\|_2
\qquad \forall\,x,y\in\mathbb{R}^{n_0}.
\]
Consequently, LipSDP computes
\begin{equation*}
\rho^*=\min_{\rho,T}\ \rho
\quad\text{subject to}\quad
\mathcal{M}(\rho,T)\preceq0,\quad T\in\mathcal{T},
\end{equation*}
and \(\sqrt{\rho^*}\) is a certified upper bound on the \(\ell_2\)-Lipschitz constant of \(f\).
\end{proposition}

The multiplier \(T\) encodes incremental quadratic constraints for the activation functions, while the block matrices \(A\), \(B\), and \(C\) encode the layer-wise network structure.
The SDP variable is \(\rho\); the resulting Lipschitz bound is \(\sqrt{\rho^*}\), not \(\rho^*\) itself.

\subsection{Lipschitz Estimation in Quantum Variational Circuits}

Quantum variational circuits (QVCs) serve as the quantum analogue of classical neural networks. In a QVC, the function is defined on the space of density operators \(\mathcal{D}(\mathcal{H})\) over a finite-dimensional Hilbert space \(\mathcal{H}\), and the output is typically a probability distribution over measurement outcomes. Recent work has extended the concept of Lipschitz continuity to QVCs, offering bounds on how output probability distributions vary with changes in the input quantum state.

\begin{lemma}[Lemma 1, \cite{guan2022fairness}]
\label{lem:existenceQlipConst}
Let \(\mathcal{C}=(\mathcal{E}, \{M_i\}_{i\in \mathcal{O}})\) be a quantum decision model, where \(\mathcal{E}\) is a CPTP map and the measurement operators satisfy \(\sum_{i\in\mathcal{O}}M_i^\dagger M_i=I\).
Then there exists a constant \( K \in (0,1] \) such that for all \(\rho,\sigma \in \mathcal{D}(\mathcal{H})\),
\[
d\big(\mathcal{C}(\rho),\mathcal{C}(\sigma)\big) \le K\, D(\rho,\sigma),
\]
where \( D(\cdot,\cdot) \) denotes the trace distance between quantum states and \( d(\cdot,\cdot) \) the total variation distance between the corresponding probability distributions.
\end{lemma}

This result, provides a theoretical guarantee on the stability of quantum decision models under state perturbations and thus a bound on the search space of $K^*$. The proof leverages the spectral properties of density operators and the structure of quantum measurements, drawing parallels with the classical analysis.

\begin{theorem}
\label{thm:lipconstant}
Let $\mathcal{C}=(\mathcal{E}, \{M_i\}_{i\in\mathcal{O}})$ be a quantum decision model and let $K^*$ be its Lipschitz constant with respect to trace distance at the input and total variation distance at the output.
Define
\[
F_i=M_i^\dagger M_i,\qquad
B_i=\mathcal{E}^\dagger(F_i),\qquad
B_A=\sum_{i\in A}B_i \quad (A\subseteq\mathcal{O}).
\]
Then
\begin{equation}
\label{eq:Qlipsdp}
K^*=\max_{A\subseteq\mathcal{O}}
\bigl[\lambda_{\max}(B_A)-\lambda_{\min}(B_A)\bigr].
\end{equation}
\end{theorem}

\begin{proof}
For an input state $\rho$, the probability of outcome $i$ is
\[
p_i(\rho)=\operatorname{tr}\!\left(F_i\mathcal{E}(\rho)\right)
=\operatorname{tr}\!\left(\mathcal{E}^\dagger(F_i)\rho\right)
=\operatorname{tr}(B_i\rho).
\]
Hence, for $\Delta=\rho-\sigma$,
\[
p_i(\rho)-p_i(\sigma)=\operatorname{tr}(B_i\Delta).
\]
Using $D(\rho,\sigma)=\frac12\|\Delta\|_1$ and total variation distance,
\[
\frac{d(\mathcal{C}(\rho),\mathcal{C}(\sigma))}{D(\rho,\sigma)}
=
\frac{\sum_{i\in\mathcal{O}}|\operatorname{tr}(B_i\Delta)|}{\|\Delta\|_1}.
\]
By homogeneity and the characterization of differences of density matrices, this gives
\[
K^*=\frac12
\max_{\substack{\Delta=\Delta^\dagger,\ \operatorname{tr}(\Delta)=0\\
                  \|\Delta\|_1\le 2}}
\sum_{i\in\mathcal{O}}|\operatorname{tr}(B_i\Delta)|.
\]
Since $\sum_iF_i=I$ and $\mathcal{E}$ is trace preserving, $\sum_iB_i=\mathcal{E}^\dagger(I)=I$.
Therefore, for $a_i=\operatorname{tr}(B_i\Delta)$ we have $\sum_i a_i=\operatorname{tr}(\Delta)=0$, and thus
\[
\frac12\sum_{i\in\mathcal{O}}|a_i|
=\max_{A\subseteq\mathcal{O}}\sum_{i\in A}a_i.
\]
Consequently,
\[
K^*=\max_{A\subseteq\mathcal{O}}
\max_{\substack{\Delta=\Delta^\dagger,\ \operatorname{tr}(\Delta)=0\\
                  \|\Delta\|_1\le 2}}
\operatorname{tr}(B_A\Delta).
\]
For a fixed Hermitian matrix $B_A$, the inner maximum equals
$\lambda_{\max}(B_A)-\lambda_{\min}(B_A)$, attained by the difference of density matrices associated with eigenvectors for the largest and smallest eigenvalues.
This proves \eqref{eq:Qlipsdp}, in agreement with Theorem~3 of \cite{guan2022fairness}.
\qed
\end{proof}
Equivalently, after setting $\delta=\Delta/2$, the same constant can be written as
\begin{equation*}
\begin{aligned}
K^*=\max_{\delta}\quad
&\sum_{i\in\mathcal{O}}\left|\operatorname{tr}(B_i\delta)\right|\\
\text{subject to}\quad
&\delta=\delta^\dagger,\qquad \operatorname{tr}(\delta)=0,\qquad
\|\delta\|_1\le 1.
\end{aligned}
\end{equation*}
The feasible set is convex, but the objective is a convex function being maximized, so this expression is not a convex optimization problem.
Equation~\eqref{eq:Qlipsdp} provides an exact finite characterization through the spectral diameter of $B_A$ for each subset $A\subseteq\mathcal{O}$.
For the numerical model in Section~\ref{sec:experiments}, the quantum output is a vector of expectation values rather than a single POVM probability vector.
Let
\[
h(\rho)=\bigl(\operatorname{tr}(H_1\rho),\ldots,\operatorname{tr}(H_m\rho)\bigr),
\]
where the $H_j$ are Hermitian observables.
For each component,
\[
\left|\operatorname{tr}\bigl(H_j(\rho-\sigma)\bigr)\right|
\le
\bigl[\lambda_{\max}(H_j)-\lambda_{\min}(H_j)\bigr]D(\rho,\sigma),
\]
and hence
\begin{equation}
\label{eq:observable-vector-bound}
\|h(\rho)-h(\sigma)\|_2
\le
K_{\rm obs}D(\rho,\sigma),\qquad
K_{\rm obs}:=
\left(\sum_{j=1}^m
[\lambda_{\max}(H_j)-\lambda_{\min}(H_j)]^2\right)^{1/2}.
\end{equation}
For the $m$ local Pauli-$Z$ observables used in the experiments, including after a common unitary conjugation by the trainable variational circuit, the observables commute, each spectral diameter is $2$, and the bound in \eqref{eq:observable-vector-bound} is attained; therefore $K_{\rm obs}=2\sqrt{m}$.
The product angle encoding used in the experiments is
\[
E(x)=\bigotimes_{j=1}^m R_X(x_j)|0\rangle\!\langle0|R_X(x_j)^\dagger.
\]
For this encoding,
\begin{equation}
\label{eq:rx-encoding-bound}
D(E(x),E(y))\le \frac12\|x-y\|_2.
\end{equation}
Indeed, for the corresponding pure states the squared overlap is
$\prod_j\cos^2((x_j-y_j)/2)$, and
$1-\prod_j\cos^2((x_j-y_j)/2)\le
\sum_j\sin^2((x_j-y_j)/2)\le\|x-y\|_2^2/4$.
Combining \eqref{eq:observable-vector-bound} and \eqref{eq:rx-encoding-bound}, the quantum block used in Section~\ref{sec:experiments} has the certified Euclidean bound $K_{\rm obs}/2=\sqrt{m}$; for four qubits this factor is $2$.

\subsection{Lipschitz Constant in Hybrid Quantum-Classical Networks}
Hybrid quantum-classical neural networks integrate classical layers with quantum variational circuits, thereby combining the strengths of both paradigms. For such networks, it is imperative to control the propagation of perturbations across the classical and quantum components.
We use the type-consistent composition of Definition~\ref{hybridnn},
\[
f = f_{k+1}\circ q_k\circ E_k\circ f_k\circ\cdots\circ q_1\circ E_1\circ f_1.
\]
The classical interfaces are measured in the Euclidean norm, the quantum-state interfaces in trace distance \(D\), and the quantum decision models in total variation distance \(d\).

\begin{theorem}
\label{thm:QLipSDP}
Let \(f\) be the hybrid quantum-classical network of Definition~\ref{hybridnn}.
Suppose that each classical layer has a certified Euclidean Lipschitz bound \(K_{f_i}\), each encoding map satisfies
\[
D(E_i(u),E_i(v))\le K_{E_i}\|u-v\|_2,
\]
and each quantum decision model has the trace-distance-to-total-variation Lipschitz constant \(K_{q_i}\) characterized in Theorem~\ref{thm:lipconstant}.

Then the Euclidean Lipschitz constant \(K_{\mathrm{hyb}}^*\) of the complete hybrid network satisfies
\begin{equation}
\label{eq:hybrid-product-bound}
K_{\mathrm{hyb}}^*
\le
K_{f_{k+1}}\prod_{i=1}^{k}
\bigl(2K_{q_i}K_{E_i}K_{f_i}\bigr).
\end{equation}
Thus, the right-hand side of \eqref{eq:hybrid-product-bound} is a certified upper bound obtained by composing the bounds of the individual components.
\end{theorem}

\begin{proof}
Let \(p=q_i(\rho)\) and \(p'=q_i(\sigma)\) be the two probability vectors produced by a quantum decision model.
Since total variation distance satisfies
\[
\|p-p'\|_2\le \|p-p'\|_1=2d(p,p'),
\]
Theorem~\ref{thm:lipconstant} and the Lipschitz property of \(E_i\) give
\[
\begin{aligned}
\|q_i(E_i(u))-q_i(E_i(v))\|_2
&\le 2K_{q_i}D(E_i(u),E_i(v))\\
&\le 2K_{q_i}K_{E_i}\|u-v\|_2.
\end{aligned}
\]
Hence \(q_i\circ E_i\) is Euclidean-Lipschitz with certified bound \(2K_{q_i}K_{E_i}\).
Composing these bounds with the Euclidean bounds \(K_{f_i}\) for the classical layers yields
\[
\|f(x)-f(y)\|_2
\le
K_{f_{k+1}}\prod_{i=1}^{k}
\bigl(2K_{q_i}K_{E_i}K_{f_i}\bigr)
\|x-y\|_2,
\]
which proves \eqref{eq:hybrid-product-bound}.
\qed
\end{proof}

For a hybrid block whose quantum component returns expectation values and satisfies
$\|q_i(\rho)-q_i(\sigma)\|_2\le K_{{\rm obs},i}D(\rho,\sigma)$,
the same composition argument replaces the factor $2K_{q_i}K_{E_i}$ in
\eqref{eq:hybrid-product-bound} by $K_{{\rm obs},i}K_{E_i}$.

This is the form used for the expectation-value quantum layer in Section~\ref{sec:experiments}.
Theorem~\ref{thm:QLipSDP} combines independently certified bounds for the classical layers, encoding maps, and quantum decision models into a certified bound for the complete hybrid network.
The result is a composition bound; it does not assert that the global Lipschitz constant is obtained by a single convex optimization problem or that the bound is tight.

\section{Experiments}
\label{sec:experiments}

We evaluate how the certified Lipschitz upper bound varies with the training objective, regularization strength, and training strategy. The experiments use the Iris dataset~\cite{iris_53} and a single stratified 80--20 train--test split. Clean test accuracy is reported alongside each certificate; these quantities measure different properties and should not be conflated.

\subsection{Experimental setup}

All experiments were run on Linux with Python~3.12, PyTorch~2.5.1, Qiskit~1.3.2, Qiskit Aer~0.16.1, Qiskit Machine Learning~0.8.2, CVXPY~1.6.0, and SCS~3.2.7. Standardization parameters were fitted on the training set only. The split uses \texttt{random\_state=42}; model initialization and minibatch ordering use training seed~1.

The model comprises two consecutive affine maps followed by ReLU, a four-qubit quantum layer, and a final affine map followed by softmax. The quantum layer uses product $R_X$ input encoding, trainable $R_Y$ rotations, and the expectation values of four local Pauli-$Z$ observables. Each model was trained for 30 epochs with batch size 16 using Adam with AMSGrad and a learning rate of $0.01$.

\subsection{Certificate and training procedures}

For certification, we combine the two affine maps preceding ReLU and apply the LipSDP relaxation to the resulting block. The expectation-value quantum layer is bounded using the trace-distance-to-Euclidean certificate for its commuting observables; product $R_X$ encoding contributes an Euclidean-to-trace-distance factor of $1/2$. We bound the final softmax layer by one half of the spectral norm of its affine weight matrix, then multiply the componentwise bounds to obtain the hybrid upper bound.

During Lipschitz-regularized training, gradients are computed using the differentiable proxy
\[
K_{\rm proxy}=\|W_2W_1\|_2\,\sqrt{m}\,\frac{\|W_{\rm out}\|_2}{2},
\]
which substitutes the spectral bound for the classical LipSDP value so that the regularizer is differentiable. For the reported certificate, an SCS result is accepted only if the solver returns an optimal status and the reconstructed LMI passes the numerical residual test. When accepted, we use the tighter of the verified LipSDP value and the always-valid spectral bound; otherwise, we use the spectral bound. Thus, all reported values are certified upper bounds.

The main comparison considers naive training, Lipschitz-regularized training, and PGD adversarial training, all with the squared $\ell_2$ data loss. Lipschitz regularization adds $\lambda K_{\rm proxy}$ to the data loss, with $\lambda=10^{-3}$. PGD uses an $\ell_\infty$ radius of $\epsilon=0.10$, step size $0.025$, four projected steps, and a random start; adversarial examples are clipped to the per-feature range of the standardized training data. In separate experiments, naive training uses each of the $\ell_1$, squared $\ell_2$, and $\ell_\infty$ losses. We also evaluate a 20-point linearly spaced sweep of $\lambda\in[10^{-4},0.5]$ for each of these three losses. The loss and regularization studies use both the original Iris labels and labels replaced by a single class.

\subsection{Results}

\begin{figure}[h]
    \begin{subfigure}{0.49\linewidth}
        \centering
        \includegraphics[width=\linewidth]{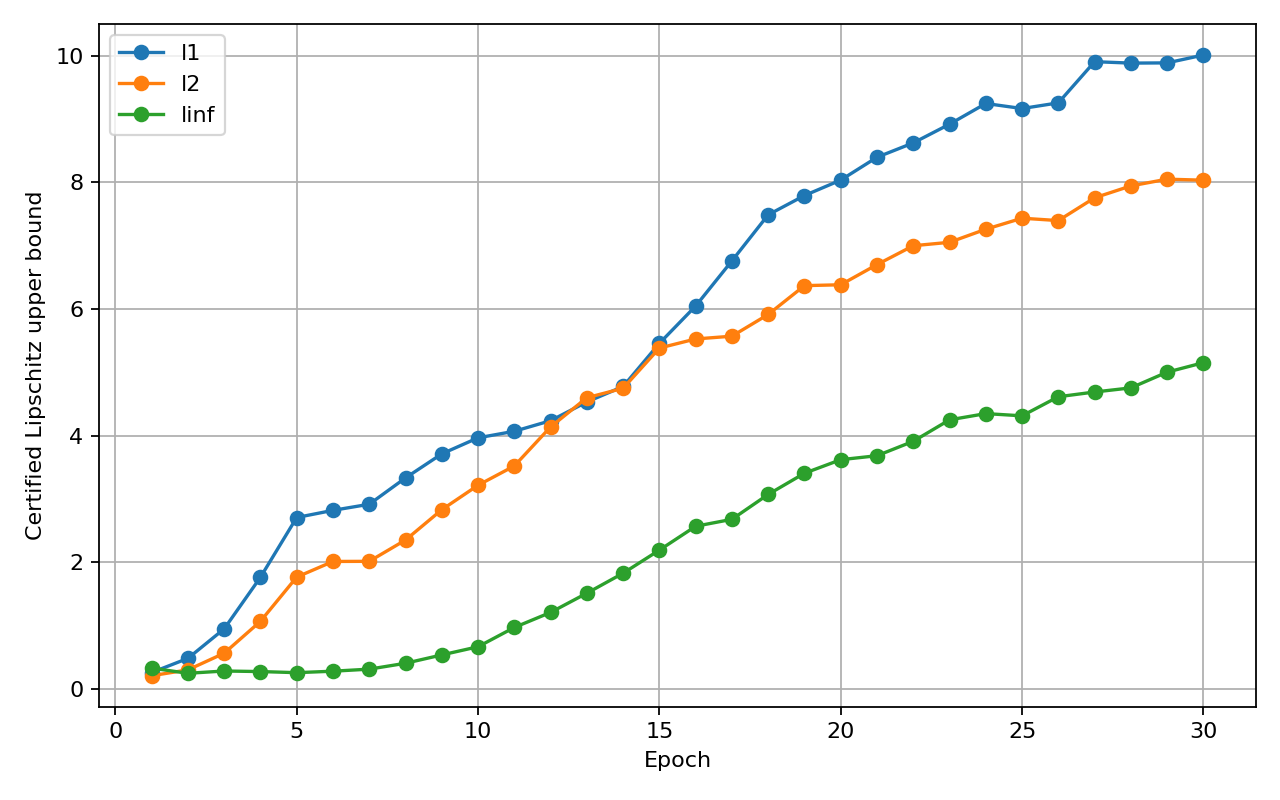}
    \end{subfigure}
    \begin{subfigure}{0.49\linewidth}
        \centering
        \includegraphics[width=\linewidth]{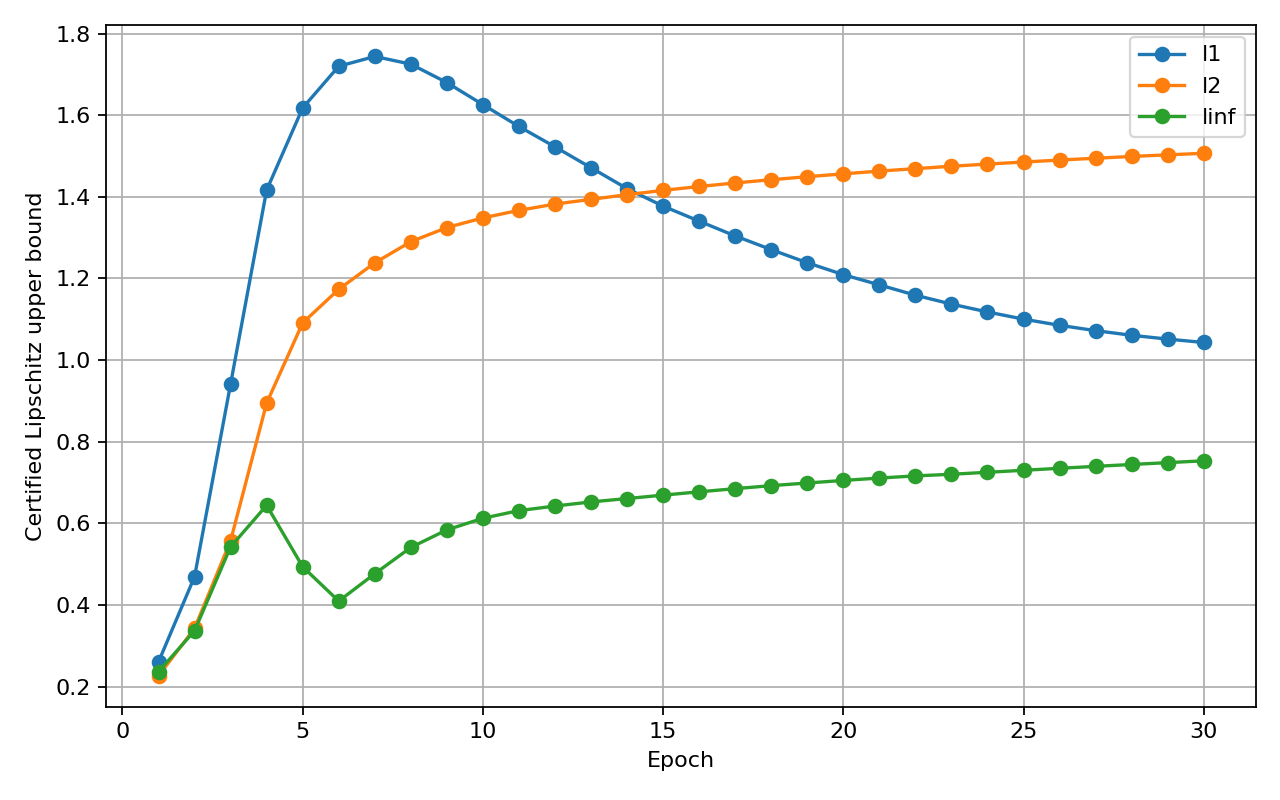}
    \end{subfigure}
    \caption{Certified Euclidean Lipschitz upper bound over 30 training epochs for naive training with $\ell_1$, squared $\ell_2$, and $\ell_\infty$ data losses. The original Iris labels are used in the left panel; all labels are replaced by a single class in the right panel. Regularization is disabled, and the certificate uses the Euclidean metric in both panels.}
    \label{fig:vsloss}
\end{figure}

\paragraph{Training loss.} Figure~\ref{fig:vsloss} shows the certified bound over training for each loss. With the original labels, the final bounds for the $\ell_1$, squared $\ell_2$, and $\ell_\infty$ losses are $10.014$, $8.036$, and $5.153$, respectively; all three runs attain 100\% clean test accuracy. When all labels are replaced by one class, the corresponding bounds are $1.043$, $1.507$, and $0.753$, again with 100\% accuracy. The loss norms describe the training objective only: every reported Lipschitz certificate uses the Euclidean metric.

\begin{figure}[h]
    \begin{subfigure}{0.49\linewidth}
        \centering
        \includegraphics[width=\linewidth]{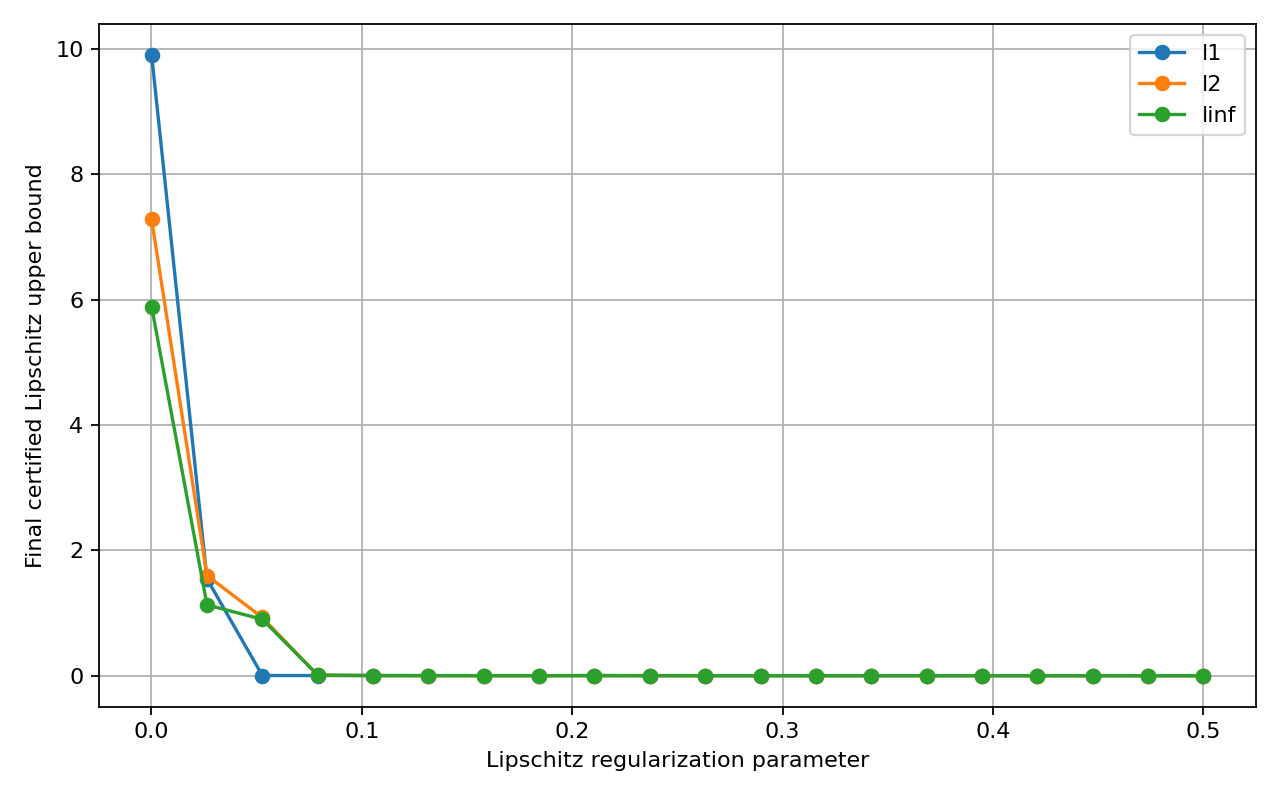}
    \end{subfigure}
    \begin{subfigure}{0.49\linewidth}
        \centering
        \includegraphics[width=\linewidth]{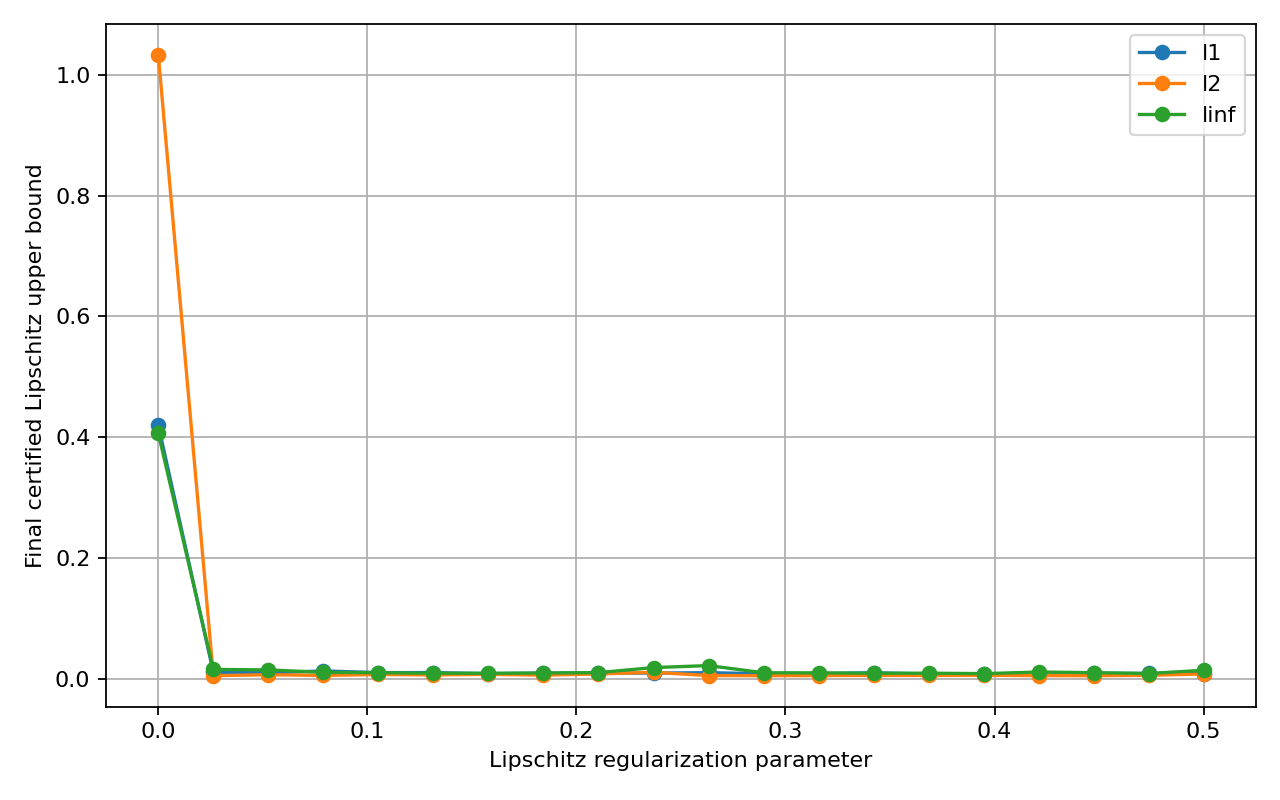}
    \end{subfigure}
    \caption{Final certified Euclidean Lipschitz upper bound as a function of the regularization coefficient for the $\ell_1$, squared $\ell_2$, and $\ell_\infty$ training losses. The original Iris labels are used in the left panel and single-class labels in the right; each panel shows the 20-point sweep over $\lambda\in[10^{-4},0.5]$. For the original labels, the reduction in the bound at larger coefficients accompanies a loss in clean test accuracy, as detailed in the text.}
    \label{fig:vsregularization}
\end{figure}

\paragraph{Regularization.} Figure~\ref{fig:vsregularization} plots the final certified bound across the regularization sweep. For the original labels and squared $\ell_2$ loss, increasing $\lambda$ from $0.0001$ to $0.0264$, $0.0527$, and $0.0790$ changes $(K,\text{clean accuracy})$ from $(7.281,100\%)$ to $(1.599,86.7\%)$, $(0.930,76.7\%)$, and $(0.0076,33.3\%)$, respectively. These results illustrate that reducing the certificate can come at the cost of predictive accuracy; the last setting is consistent with the classifier approaching a constant map. In the single-class relabeling experiment, accuracy remains 100\% because a constant classifier matches the target labels.

\begin{figure}[h]
    \begin{subfigure}{0.49\linewidth}
        \centering
        \includegraphics[width=\linewidth]{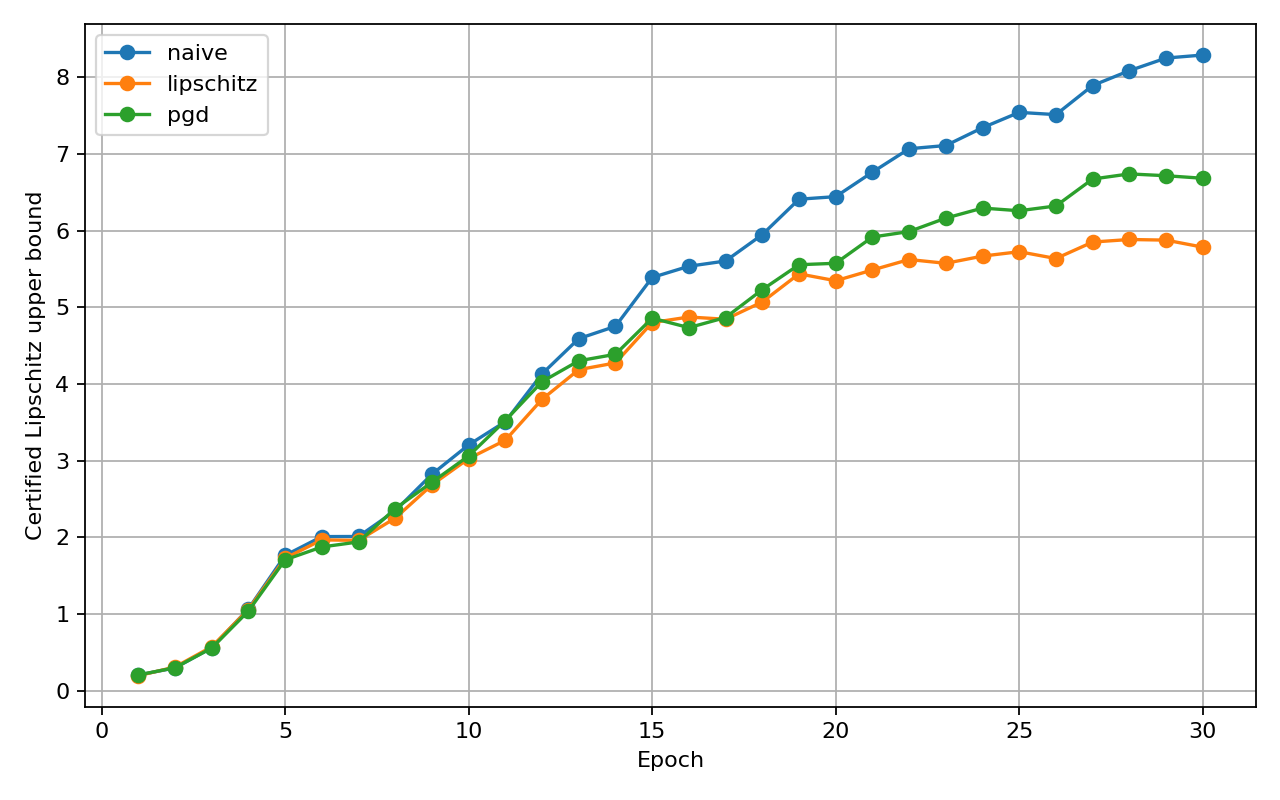}
    \end{subfigure}
    \begin{subfigure}{0.49\linewidth}
        \centering
        \includegraphics[width=\linewidth]{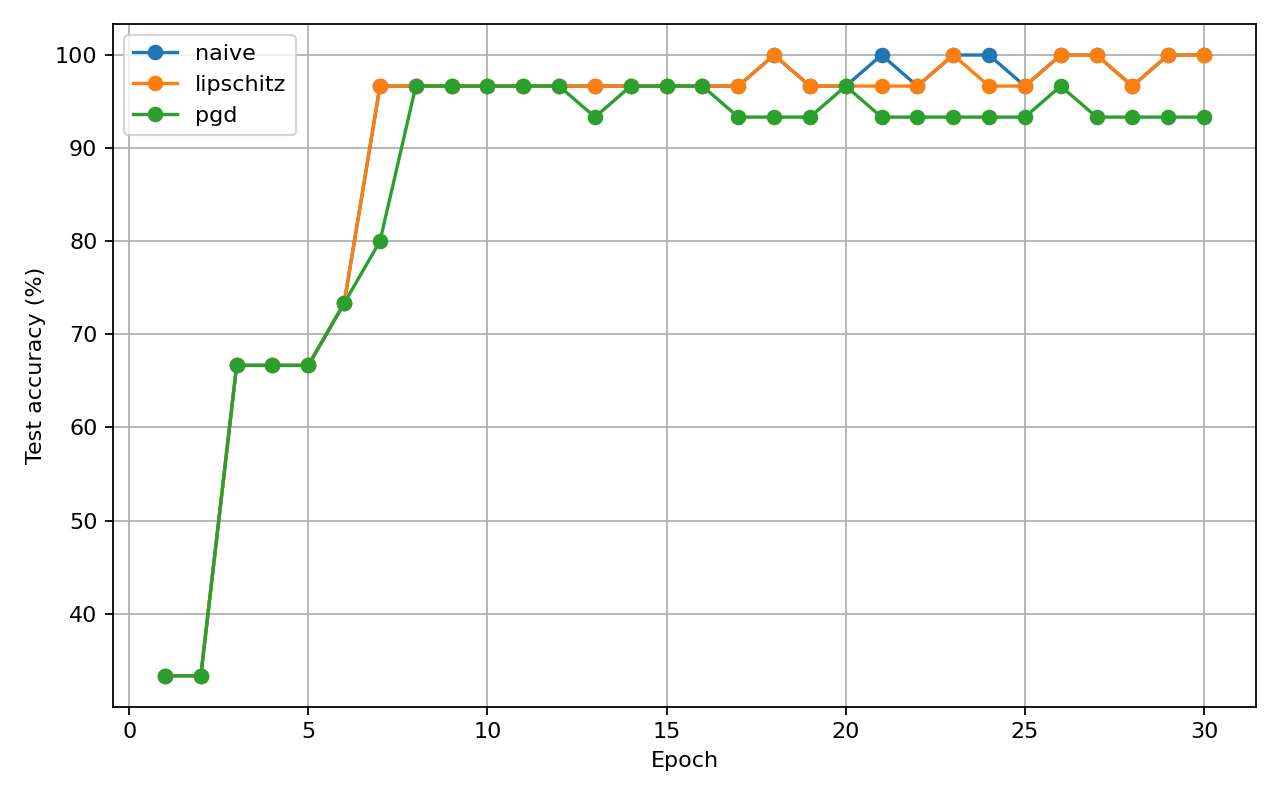}
    \end{subfigure}
    \caption{Training trajectories for naive training, Lipschitz-regularized training, and PGD adversarial training with squared $\ell_2$ data loss. The left panel reports the certified Euclidean Lipschitz upper bound; the right panel reports clean test accuracy (not adversarial accuracy). The epoch-30 values are summarized in Table~\ref{tab:training-comparison}.}
    \label{fig:vsaccuracy}
\end{figure}

\paragraph{Training strategy.} Table~\ref{tab:training-comparison} and Fig.~\ref{fig:vsaccuracy} compare the three strategies. At epoch 30, naive training yields a certified bound of $8.288$ and 100\% clean test accuracy. Lipschitz regularization yields a bound of $5.784$ with the same accuracy, a 30.2\% reduction in the bound. PGD yields a bound of $6.682$ and 93.3\% clean test accuracy. These figures report clean accuracy only and do not establish adversarial test accuracy. Because the results are based on one split of the Iris dataset, they should be interpreted as a controlled illustration rather than evidence of performance across data splits.

\begin{table}[h]
\centering
\caption{Certified Lipschitz upper bounds and clean test accuracy at epoch 30 for naive training, Lipschitz regularization, and PGD adversarial training.}
\label{tab:training-comparison}
\begin{tabular}{l@{\hspace{1.5em}}r@{\hspace{2em}}r}
\hline
Training method & Certified bound & Clean accuracy \\
\hline
Naive & 8.288 & 100.0\% \\
Lipschitz regularization & 5.784 & 100.0\% \\
PGD & 6.682 & 93.3\% \\
\hline
\end{tabular}
\end{table}

\section{Conclusion}

In this work, we developed a framework for certifying Lipschitz bounds for hybrid quantum-classical neural networks.
Our framework uses an SDP-based LipSDP certificate for classical neural-network layers and an exact spectral characterization for probability-output quantum decision models. It then combines these componentwise bounds through an explicit composition bound that accounts for the encoding maps.
The numerical section examines the relation between the reported Lipschitz estimates, training strategies, and model accuracy.
Future work can examine deeper hybrid architectures, additional quantum layers, and training schemes that adapt the strength of Lipschitz regularization during optimization.
A separate direction is to evaluate adversarial robustness directly, rather than inferring robustness from a Lipschitz certificate or clean test accuracy alone.

\section*{Acknowledgements}
We would like to thank Makoto Yamashita\footnote{Tokyo Institute of Technology, \email{makoto.yamashita@c.titech.ac.jp}} for his careful reviews and helpful feedback on this article. We also thank Ebrahim Ardeshir-Larijani \footnote{Iran University of Science and Technology, \email{larijani@iust.ac.ir}} for the guidance and answering several questions about \cite{guan2022fairness}.

%
%
\bibliographystyle{splncs04}
\bibliography{bib}

\end{document}